\documentclass[superscriptaddress]{revtex4-1} 
\usepackage{revsymb4-1}
\usepackage{amssymb}
\usepackage{amsmath}
\usepackage{graphicx}
\usepackage{dcolumn}
\usepackage{bm} 
\usepackage{caption}
\usepackage{subcaption}
\begin{document}
\title{Polaronic effects in monolayer black phosphorus on polar substrates}
\author{A. Mogulkoc}
\email{A.Mogulkoc@science.ru.nl}
\email{mogulkoc@science.ankara.edu.tr}
\affiliation{Radboud University, Institute for Molecules and Materials, Heyendaalseweg 135, 6525 AJ Nijmegen, The Netherlands}
\affiliation{Department of Physics, Faculty of Sciences, Ankara University, 06100\\
Tandogan, Ankara, Turkey}
\author{Y. Mogulkoc}
\affiliation{Radboud University, Institute for Molecules and Materials, Heyendaalseweg 135, 6525 AJ Nijmegen, The Netherlands}
\affiliation{Department of Engineering Physics, Faculty of Engineering, Ankara University, 06100\\
	Tandogan, Ankara, Turkey}
\author{A. N. Rudenko}
\affiliation{Radboud University, Institute for Molecules and Materials, Heyendaalseweg 135, 6525 AJ Nijmegen, The Netherlands}
\author{M. I. Katsnelson}
\affiliation{Radboud University, Institute for Molecules and Materials, Heyendaalseweg 135, 6525 AJ Nijmegen, The Netherlands}
\date{\today}

\begin{abstract}
We investigate the effect of charge carrier interaction with surface optical phonons on the band properties of monolayer black phosphorus induced by polar substrates. We develop an analytical method based on the Lee-Low-Pines theory to calculate the spectrum of Fr{\"{o}}hlich type continuum Hamiltonian in the long-wavelength limit. We examine the modification of a band gap and renormalization of effective masses due to the substrate-related polaronic effect. Our results show that an energy gap in supported monolayer black phosphorus is enlarged depending on a particular substrate and the interlayer distance, $z$. Among the substrate considered, the largest gap broadening at $ z=2.5$ \AA{} is observed for the Al$ _{2} $O$ _{3} $ substrate, which is found to be $ \sim 50$ meV. Carrier-phonon coupling also renormalizes the effective masses which is more pronounced along the zigzag direction. Anisotropy of the effective masses becomes stronger by the influence of the polaronic effect corresponding to direction-dependent carrier-phonon coupling. We conclude that substrate phonons have a non-negligible effect on the static band properties of monolayer black phosphorus, which may be further exploited in its experimental and theoretical studies.
\end{abstract}
\maketitle

\section{Introduction}

With the discovery of graphene \cite{novoselov2004electric}, two-dimensional (2D) materials have become the focus of many researchers. Recently, a few-layer black phosphorus (BP) has emerged as one of the promising 2D material owing to its unique physical properties. Its peculiar structure leads significantly anisotropic electronic and optical properties \cite{Li:2014aa,doi:10.1021/nn501226z,Xia:2014aa,Reich2014,koenig2014electric,castellanos2014isolation,Low:2014aa}. Similar to graphene, BP can be mechanically exfoliated \cite{Li:2014aa,doi:10.1021/nn501226z} down to a few layers sample. Beside this, few-layer BP has been obtained using liquid-phase exfoliation \cite{brent2014production,yasaei2015high} and plasma-assisted fabrication method \cite{lu2014plasma}. Chemically, BP is less stable than graphene and it degrades quickly in the atmospheric environment \cite{island2015environmental}. To protect BP samples from the degradation, their encapsulation in other materials is used to achieve better performance of BP-based devices. For example, high carrier mobility of $ \sim $1.350 $ cm^{2}V^{-1}s^{-1} $ at room temperature and high on-off ratios exceeding $ \sim10^{5} $ in a few layer BP encapsulated by atomically thin hexagonal boron nitride (h-BN), which forms a BN-BP-BN heterostructures is demonstrated by Chen \textit{et al.} \cite{Chen:2015aa}. Interaction with surfaces is, therefore, an important aspect affecting the properties of BP in its practical applications.

 Lattice vibrations play an important role in the dynamics of charge carriers in 2D materials \cite{Samsonidze:2007aa,Stauber:2007aa,Hague:2011aa,Hague:2012aa,Hague:2014aa,Kandemir:2012aa,kandemir2014zone,wang2015energy,kandemir2015chiral,Mogulkoc:2015aa,mogulkoc2015ab}. Particularly, optical phonons of polar substrates localized around sample-substrate interface affect the behavior of charge carrier, which depends strongly on the phonon frequencies and polarizability of a substrate \cite{Chen:2008aa,Fratini:2008aa,konar2010effect,Hague:2011aa,Zhang:2013aa,Hwang:2013aa,Tan2013aa,Scharf:2013aa,wang2015energy}. Similar to the effect of strain \cite{Peng:2014aa,doi:10.1021/nl500935z}, interaction of electrons with optical phonons can open a band gap or allow us to tune the existing band gap of the system. Band gap engineering is an important field aiming at tuning on energy gap of material for potential applications in nanoelectronics. For instance, the electron-phonon interaction can induce a small gap in the band structure of graphene \cite{Samsonidze:2007aa,Hague:2011aa,Hague:2012aa,Hague:2014aa,kandemir2014zone,wang2015energy,kandemir2015chiral}. Moreover, Wang \textit{et al.} \cite{wang2015energy} found an energy gap in the zeroth Landau level due to the electron-surface optical phonon interaction arising from the polar substrate, this gap can be tuned by choosing the polarization of substrates and changing the distance between the substrate and graphene. Based on density functional theory calculations, Hu and Hong \cite{hu2015anisotropic} have recently found that the encapsulation of BP layers into h-BN changes the band structure affecting the band gap and effective masses without changing the anisotropic electrical and optical properties of BP layers. Such modification is a result of a single-particle approximation, while the role of the many-body effects remains unclear. 
 
 Here, we investigate another mechanism responsible for the formation of band properties of encapsulated or supported monolayer black phosphorus (MBP), namely the surface polaronic effect. We consider an analytical model based on the Lee-Low-Pines (LLP) theory \cite{lee1953motion} which is based on a $ \boldsymbol{k}\cdot \boldsymbol{p} $-type Hamiltonian for MBP. We apply a series of unitary transformations to diagonalize the Fr{\"{o}}hlich type many-body Hamiltonian and use the variational method to find the eigenvalues of the total Hamiltonian. The band structure of MBP is then investigated as a function of model parameters, such as phonon energy, $ \hbar \omega $, effective screening constant, $ \xi $, and the distance between substrate and MBP, $ z $. Finally, the gap modifications and renormalization of the effective masses are examined by considering a number of typical polar substrates (h-BN, HfO$ _{2} $, SiC, SiO$ _{2} $, Al$ _{2} $O$ _{3} $).    
\section{Theory}

\begin{figure} [h]
    \centering
    \includegraphics[width=0.45\textwidth]{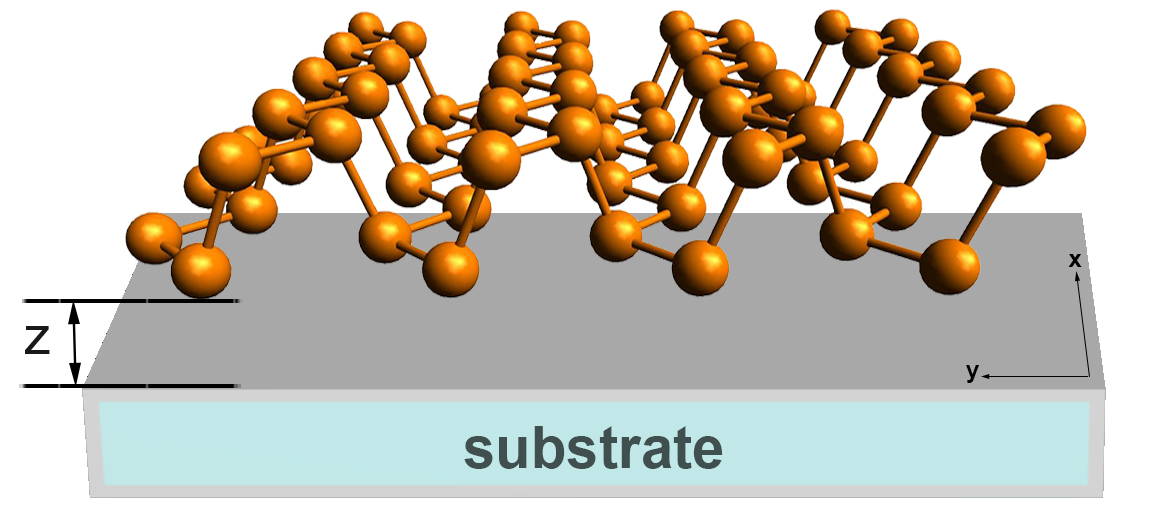}
    \caption{Illustration of a MBP/substrate system. $ z $ is the vertical distance between MBP and a polar substrate.}
    \label{FIGURE1}
\end{figure}

The energy spectrum of MBP can be described by a tight-binding model \cite{Rudenko:2014aa,Rudenko:2015aa} and effective $ \boldsymbol{k}\cdot \boldsymbol{p} $ model \cite{Rodin:2014aa,1504.02452v1,Zhou:2015aa,ezawa2014topological} as a continuum approach, which shows good agreement with the tight-binding results in the range of $ -2.0$ to $1.5 $ eV \cite{1504.02452v1,Zhou:2015aa}. The band structure of MBP exhibits a direct band gap of $ 1.5 \sim 2.0 $ eV at the $ \Gamma $ point of the Brillouin zone \cite{doi:10.1021/nn501226z,Rudenko:2014aa,Rudenko:2015aa}. In the long-wavelength regime, the continuum Hamiltonian of MBP \cite{1504.02452v1} with bare phonon field and electron-surface optical phonon Hamiltonian can be written as, 
\begin{equation}
\mathcal{H}=\mathcal{H}_{0}+\sum\limits_{\mu }\sum\limits_{\boldsymbol{q}%
}\hbar \omega _{\mu }b_{\boldsymbol{q}}^{\dag }b_{%
\boldsymbol{q}}+\mathcal{H}_{e-p}  \label{1}
\end{equation}%
with
\begin{equation}
\mathcal{H}_{0}=\left[ 
\begin{array}{cc}
u_{0}+\bar{\eta}_{x}p_{x}^{2}+\bar{\eta}_{y}p_{y}^{2} & \delta+\bar{\gamma}_{x}p_{x}^{2}+\bar{\gamma}_{y}p_{y}^{2}+i\bar{\chi}p_{y} \\ 
\delta+\bar{\gamma}_{x}p_{x}^{2}+\bar{\gamma}_{y}p_{y}^{2}-i\bar{\chi}p_{y} & u_{0}+\bar{\eta}_{x}p_{x}^{2}+\bar{\eta}_{y}p_{y}^{2}
\end{array}%
\right]  \label{2}
\end{equation}%
and
\begin{equation}
\mathcal{H}_{e-p}=\sum\limits_{\mu}\sum\limits_{\boldsymbol{q}}\left[ \mathcal{M}_{\mu \boldsymbol{q}}b_{\boldsymbol{q}}e^{i\boldsymbol{q}\cdot\boldsymbol{r}}+h.c.\right], \label{3}
\end{equation}%
where $\mathcal{M}_{\mu\boldsymbol{q}}=\sqrt{e^{2}\xi\hbar\omega_{\mu}/2A\epsilon_{0}q}e^{-qz}$ is the amplitude of carrier-surface optical phonon interaction \cite{Wang:1972aa,Mori:1989aa,Sak:1972aa}, $ b_{\boldsymbol{k}}^{\dag }(b_{\boldsymbol{k}}) $ is the phonon creation (annihilation) operator, $ e $ is the elementary electric charge, $ \xi =\left( \kappa_{0}-\kappa_{\infty}\right)/\left[\left(\kappa_{0}+1 \right)\left(\kappa_{\infty}+1 \right)  \right]  $ is the effective screening constant related to the dielectric constants of a substrate, $ \omega_{\mu} $ is the longitudinal surface optical phonon frequency of the $ \mu^{th} $ branch ($ \mu=1,2 $), $ z $ is the distance between the substrate and MBP [see Fig.(\ref{FIGURE1})], $ \kappa_{\infty} $ ($ \kappa_{0} $) is the high (low) frequency dielectric constant of the substrate and $ \epsilon_{0} $ is the vacuum permittivity. Here, $ \bar{\eta}_i=\eta_{i}/\hbar^{2} $ ($ \eta_{x}=0.58 $ eV\AA{}$ ^{2} $ and $ \eta_{y}=1.01 $ eV\AA{}$ ^{2} $), $ \bar{\gamma}_i=\gamma_{i}/\hbar^{2} $ ($ \gamma_{x}=3.93 $ eV\AA{}$ ^{2} $ and $ \gamma_{y}=3.83 $ eV\AA{}$ ^{2} $), $ \bar{\chi}=\chi/\hbar $ ($ \chi=5.25 $ eV\AA{}), $ u_{0}=-0.42 $ eV and $ \delta=0.76 $ eV. The total Hamiltonian given by Eq.(\ref{1}) can be rewritten in a more compact form as, 

\begin{eqnarray}
\mathcal{H}&=&\left( u_{0}+\sum\limits_{i=x,y}\bar{\eta}_{i}p_{i}^{2}\right)\sigma_{0}+\left(\delta +\sum\limits_{i=x,y}\bar{\gamma}_{i}p_{i}^{2}\right)\sigma_{1}-\left( \bar{\chi} p_{y} \right)\sigma_{2} \notag \\
&+& \left(\ \sum\limits_{\boldsymbol{q} \mu}\mathcal{M}_{\mu \boldsymbol{q}}b_{\boldsymbol{q}}e^{i\boldsymbol{q}\cdot\boldsymbol{r}}+h.c.\right)\sigma_{0}+\left( \sum\limits_{\mu }\sum\limits_{\boldsymbol{q}%
}\hbar \omega _{\mu }b_{\boldsymbol{q}}^{\dag }b_{%
\boldsymbol{q}}\right) \sigma_{0} \label{4}
\end{eqnarray}

Here, $ \sigma_{i} $ $ \left(i=1,2,3 \right)  $ are the Pauli matrices, and $ \sigma_{0} $ is the identity matrix. Note that, electron-phonon interaction and bare phonon field are involved in the low-energy Hamiltonian with identity matrix which equally affects both sublattices. On the other hand, we assume that the substrate is isotropic and the interlayer interaction is weak which preserves C$_{2h}$ group invariance of the supported MBP lattice. To diagonalize the Hamiltonian in Eq.(\ref{4}), we introduce two unitary transformation given by
\begin{eqnarray}
U_{1}&=&\exp \left[ -i\boldsymbol{r}\mathbf{\cdot }\sum\limits_{%
	\boldsymbol{q} \mu}\boldsymbol{q} b_{\boldsymbol{q}%
}^{\dag }b_{\boldsymbol{q}}\right] \notag \\
U_{2}&=&\exp \left[\sum\limits_{\boldsymbol{q}%
}\left(\textit{f}_{\boldsymbol{q}} b_{\boldsymbol{q}}^{\dag}-\textit{f}_{\boldsymbol{q}}^{*} b_{\boldsymbol{q}}\right)  \right].\notag
\end{eqnarray}%

The first transformation eliminates the electron coordinates from Eq.(\ref{4}) and transforms the phononic and momentum operators by $ b_{\boldsymbol{q}}\rightarrow b_{\boldsymbol{q}}\exp\left(i \boldsymbol{q} \cdot \boldsymbol{r} \right) $ and $ p_{i}\rightarrow p_{i}-\sum_{\boldsymbol{q}} \hbar \boldsymbol{q}b_{\boldsymbol{q}}^{\dag }b_{%
\boldsymbol{q}}$, respectively. The second transformation is the well-known LLP transformation \cite{lee1953motion}, where $ \textit{f}_{\boldsymbol{q}} $ is considered to be a variational function. This transformation enables us to consider the dressed electron states due to the phonon field induced by the substrate which yields the polarization in the lattice. Moreover, it generates the coherent states for the phonon subsystem such that optical phonon operators transform according to the rule $ b_{\boldsymbol{q}}\rightarrow b_{\boldsymbol{q}}+\textit{f}_{q} $. Under the two successive unitary transformation $ \widetilde{\mathcal{H}}=U_{2}^{-1}U_{1}^{-1}\mathcal{H}U_{1}U_{2} $, the transformed Hamiltonian takes the form
\begin{eqnarray}
\widetilde{\mathcal{H}}&=&\left[u_{0}+\sum\limits_{i=x,y}\bar{\eta}_{i}p_{i}^{2} +\sum\limits_{i=x,y} \bar{\eta}_{i} \left( \sum\limits_{\boldsymbol{q}}\hbar q_{i} \left|\textit{f}_{\boldsymbol{q}} \right|^{2} \right)^{2}+\sum\limits_{\boldsymbol{q} \mu} \left( \mathcal{M}_{\mu \boldsymbol{q}} \textit{f}_{\boldsymbol{q}}+\mathcal{M}_{\mu \boldsymbol{q}}^{*} \textit{f}_{\boldsymbol{q}}^{*}\right)  \right]\sigma_{0} \notag \\
&+& \left[ \sum\limits_{\boldsymbol{q}}\left|\textit{f}_{\boldsymbol{q}} \right|^{2} \left( \hbar \omega_{\mu}-2\hbar\sum\limits_{i=x,y}\bar{\eta}_{i} p_{i}q_{i}+\hbar^{2}\sum\limits_{i=x,y}\bar{\eta}_{i}q_{i}^{2}\right) \right] \sigma_{0}+\left[\delta+\sum\limits_{i=x,y}\bar{\gamma}_{i}p_{i}^{2} +\sum\limits_{i=x,y} \bar{\gamma}_{i} \left( \sum\limits_{\boldsymbol{q}}\hbar q_{i} \left|\textit{f}_{\boldsymbol{q}} \right|^{2} \right)^{2} \right] \sigma_{1} \notag \\
&+&\left[ \sum\limits_{\boldsymbol{q}}\left|\textit{f}_{\boldsymbol{q} } \right|^{2} \left(-2\hbar\sum\limits_{i=x,y}\bar{\gamma}_{i} p_{i}q_{i}+\hbar^{2}\sum\limits_{i=x,y}\bar{\gamma}_{i}q_{i}^{2}\right) \right] \sigma_{1}-\bar{\chi}\left(p_{y}- \sum\limits_{\boldsymbol{q}}\hbar q_{y} \left|\textit{f}_{\boldsymbol{q}} \right|^{2}\right)\sigma_{2}+\left( \sum\limits_{\boldsymbol{q} \mu%
}\hbar \omega _{\mu }b_{\boldsymbol{q}}^{\dag }b_{%
\boldsymbol{q}}\right) \sigma_{0} \notag \\
&+&\sum\limits_{\boldsymbol{q}%
} \left[ \left( \sum\limits_{i=x,y} 2\hbar^{2} \bar{\eta}_{i}q_{i}\sum\limits_{\boldsymbol{q^{'}}} q_{i}^{'} \left|\textit{f}_{\boldsymbol{q}^{'}} \right|^{2}\right) \sigma_{0}+ \left( \sum\limits_{i=x,y} 2\hbar^{2} \bar{\gamma}_{i}q_{i}\sum\limits_{\boldsymbol{q^{'}}} q_{i}^{'} \left|\textit{f}_{\boldsymbol{q}^{'}} \right|^{2}\right) \sigma_{1} \right]b_{\boldsymbol{q}}^{\dag }b_{%
\boldsymbol{q}}+\mathcal{\widetilde{H}}_{ND},  \label{5}
\end{eqnarray}
where $ \mathcal{\widetilde{H}}_{ND} $ includes nondiagonal elements in terms of phonon creation and annihilation operators. It is also convenient to express the kinetic energy and linear momentum operators in the second quantization form as,
\begin{eqnarray}
\bar{\eta}_{i}p_{i}^{2}&=&\hbar\sum\limits_{\boldsymbol{k}}\eta_{i} k_{i}^{2}a_{\boldsymbol{k}}^{\dag }a_{%
	\boldsymbol{k}} \notag \\
\bar{\gamma}_{i}p_{i}^{2}&=&\hbar\sum\limits_{\boldsymbol{k}}\gamma_{i} k_{i}^{2}a_{\boldsymbol{k}}^{\dag }a_{%
	\boldsymbol{k}}\notag \\
p_{i}&=&\hbar\sum\limits_{\boldsymbol{k}}k_{i}a_{\boldsymbol{k}}^{\dag }a_{%
	\boldsymbol{k}} \notag
\end{eqnarray}
with the following vacuum state vector,
\begin{eqnarray}
\left| k_{i}, 0\right\rangle &=&a_{\boldsymbol{k}}^{\dag }\left| 0,0\right\rangle \notag \\
\left|0,0 \right\rangle &=& \left|0 \right\rangle_{e}\left| 0\right\rangle_{ph},  \notag 
\end{eqnarray}
where $ \left|0 \right\rangle_{e} $ ($ \left| 0\right\rangle_{ph} $) is the vacuum state vector for electron (phonon), and $ a_{\boldsymbol{k}}^{\dag }(a_{\boldsymbol{k}}) $ is the electron creation (annihilation) operator. Using the vacuum state vectors, the expectation values of the transformed Hamiltonian ($ \widetilde{H}^{'}=\left\langle 0,0\right| \widetilde{H}\left| 0,0 \right\rangle  $) can be written as
\begin{eqnarray}
\widetilde{H}^{'}&=&\left[u_{0}+\sum\limits_{i=x,y}\bar{\eta}_{i}k_{i}^{2} +\sum\limits_{i=x,y} \bar{\eta}_{i} \left( \sum\limits_{\boldsymbol{q}}\hbar q_{i} \left|\textit{f}_{\boldsymbol{q}} \right|^{2} \right)^{2}+\sum\limits_{\boldsymbol{q} \mu} \left( \mathcal{M}_{\mu \boldsymbol{q}} \textit{f}_{\boldsymbol{q}}+\mathcal{M}_{\mu \boldsymbol{q}}^{*} \textit{f}_{\boldsymbol{q}}^{*}\right)  \right]\sigma_{0} \notag \\
&+&\left[ \sum\limits_{\boldsymbol{q}}\left|\textit{f}_{\boldsymbol{q}} \right|^{2} \left( \hbar \omega_{\mu}-2\hbar\sum\limits_{i=x,y}\bar{\eta}_{i} k_{i}q_{i}+\hbar^{2}\sum\limits_{i=x,y}\bar{\eta}_{i}q_{i}^{2}\right) \right] \sigma_{0}+\left[\delta+\sum\limits_{i=x,y}\bar{\gamma}_{i}k_{i}^{2} +\sum\limits_{i=x,y} \bar{\gamma}_{i} \left( \sum\limits_{\boldsymbol{q}}\hbar q_{i} \left|\textit{f}_{\boldsymbol{q}} \right|^{2} \right)^{2} \right] \sigma_{1} \notag \\
&+& \left[ \sum\limits_{\boldsymbol{q}}\left|\textit{f}_{\boldsymbol{q}} \right|^{2} \left(-2\hbar\sum\limits_{i=x,y}\bar{\gamma}_{i} k_{i}q_{i}+\hbar^{2}\sum\limits_{i=x,y}\bar{\gamma}_{i}q_{i}^{2}\right) \right] \sigma_{1}-\bar{\chi}\left(k_{y}- \sum\limits_{\boldsymbol{q}}\hbar q_{y} \left|\textit{f}_{\boldsymbol{q}} \right|^{2}\right)\sigma_{2} \label{6}
\end{eqnarray}

Eq.(\ref{6}) can be minimized by employing the following conditions,
\begin{equation}
\dfrac{\delta\widetilde{H}^{'}}{\delta \textit{f}_{\boldsymbol{q}}}=0, \quad \dfrac{\delta\widetilde{H}^{'}}{\delta \textit{f}_{\boldsymbol{q}}}^{*}=0. \notag
\end{equation}
yielding the equation on $ \textit{f}_{\boldsymbol{q}} $,

\begin{eqnarray}
&&\left[ \mathcal{M}_{\mu \boldsymbol{q}}+\textit{f}_{\boldsymbol{q}}^{*} \left(\hbar \omega_{\mu}-2\sum\limits_{i=x,y}{\eta}_{i} k_{i}q_{i}+\sum\limits_{i=x,y}{\eta}_{i}q_{i}^{2}+2\sum\limits_{\boldsymbol{q^{'}}}{\eta}_{i} \left( q_{i}^{'} \left|\textit{f}_{\boldsymbol{q}^{'}} \right|^{2} \right) ^{2}\right) \right] \sigma_{0} \notag \\
&+&\left[\textit{f}_{\boldsymbol{q}}^{*} \left(-2\sum\limits_{i=x,y}{\gamma}_{i} k_{i}q_{i}+\sum\limits_{i=x,y}{\gamma}_{i}q_{i}^{2}+2\sum\limits_{\boldsymbol{q^{'}}}{\gamma}_{i} \left( q_{i}^{'} \left|\textit{f}_{\boldsymbol{q}^{'}} \right|^{2} \right) ^{2}\right) \right] \sigma_{1}+\left( \textit{f}_{\boldsymbol{q}}^{*} \chi q_{y}\right) \sigma_{2}=0. \label{7} 
\end{eqnarray}
The only preferred direction in this problem is chosen that of $ \boldsymbol{p} $ following the original work of LLP \cite{lee1953motion}. Thus, the new variation parameters can be introduced as,
\begin{equation}
\alpha_{i}k_{i}=\sum\limits_{\boldsymbol{q}}q_{i} \left|\textit{f}_{\boldsymbol{q}} \right|^{2} \label{8}
\end{equation}
Here, $ \alpha_{x} $ and $ \alpha_{y} $ are the new variation parameters along the $ x $ and $ y $ directions, respectively. $ \alpha_{i} $ is a fraction ($ \alpha_{i}<1 $) which is valid for weak and intermediate coupling regime. Note that in Ref.\onlinecite{lee1953motion}, Eq.(\ref{8}) was evaluated analytically by expanding the expression to order of $ ~\boldsymbol{p} $. It is also obvious that, $ \alpha_{i} $ can be regarded as variation parameters, which minimize the energy eigenvalues related with the coupling strength of interaction (for further discussion of the LLP method, see Ref.\onlinecite{singh2013excitation,1012.4576v5}). The minimum of the quadratic expression in Eq.(\ref{7}) can be found by using the variation parameters, $ \textit{f}_{\boldsymbol{q}}^{*} $ as,
\begin{equation}
\textit{f}_{\boldsymbol{q}}^{*}=\dfrac{-\mathcal{M}_{\mu \boldsymbol{q}}}{\mathbb{I}_{1}\left(\boldsymbol{k},\boldsymbol{q},\boldsymbol{\alpha},\mu \right) +\left(\mathbb{I}_{2}^{2}\left(\boldsymbol{k},\boldsymbol{q},\boldsymbol{\alpha}\right)+\left(\chi q_{y} \right)^{2}  \right)^{1/2} } \label{9}
\end{equation} 
where
\begin{eqnarray}
\mathbb{I}_{1}\left(\boldsymbol{k},\boldsymbol{q},\boldsymbol{\alpha},\mu  \right)&=&\hbar \omega_{\mu}-2\sum\limits_{i=x,y}\left(1-\alpha_{i} \right) {\eta}_{i} k_{i}q_{i}+\sum\limits_{i=x,y}{\eta}_{i}q_{i}^{2} \notag \\
\mathbb{I}_{2}\left(\boldsymbol{k},\boldsymbol{q},\boldsymbol{\alpha} \right)&=&-2\sum\limits_{i=x,y}\left(1-\alpha_{i} \right) {\gamma}_{i} k_{i}q_{i}+\sum\limits_{i=x,y}{\gamma}_{i}q_{i}^{2} \notag
\end{eqnarray}
Using Eqs.(\ref{8}) and (\ref{9}), the eigenvalues of the Hamiltonian given by Eq.(\ref{6}) can be found as,
\begin{eqnarray}
E_{\lambda}&=&\left(u_{0}+\sum\limits_{i=x,y}\left( \alpha_{i}^{2}+1\right) \eta k_{i}^{2}  \right) \left\langle \sigma_{0}\right\rangle+\sum\limits_{\boldsymbol{q} \mu} \left( \frac{-2\left|\mathcal{M}_{\mu \boldsymbol{q}} \right|^{2} }{\mathbb{I}_{1}\left(\boldsymbol{k},\boldsymbol{q},\boldsymbol{\alpha},\mu  \right) +\left(\mathbb{I}_{2}^{2}\left(\boldsymbol{k},\boldsymbol{q},\boldsymbol{\alpha} \right)+\left(\chi q_{y} \right)^{2}  \right)^{1/2} }\right) \left\langle \sigma_{0}\right\rangle \notag \\
&+&  \sum\limits_{\boldsymbol{q} \mu}\left(  \frac{\left|\mathcal{M}_{\mu \boldsymbol{q}} \right|^{2} \left( \mathbb{I}_{1}\left(\boldsymbol{k},\boldsymbol{q},\boldsymbol{\alpha},\mu  \right)-2\sum\limits_{i=x,y}\alpha_{i}\eta_{i}k_{i}^{2}\right) }{\left( \mathbb{I}_{1}\left(\boldsymbol{k},\boldsymbol{q},\boldsymbol{\alpha},\mu  \right) +\left(\mathbb{I}_{2}^{2}\left(\boldsymbol{k},\boldsymbol{q},\boldsymbol{\alpha} \right)+\left(\chi q_{y} \right)^{2}  \right)^{1/2} \right)^{2} }\right) \left\langle \sigma_{0}\right\rangle  +\left(\delta+\sum\limits_{i=x,y}\left( \alpha_{i}^{2}+1\right) \gamma k_{i}^{2}  \right) \left\langle \sigma_{1}\right\rangle  \notag \\
&+&  \sum\limits_{\boldsymbol{q} \mu}\left(  \frac{\left|\mathcal{M}_{\mu \boldsymbol{q}} \right|^{2} \left( \mathbb{I}_{2}\left(\boldsymbol{k},\boldsymbol{q},\boldsymbol{\alpha}  \right)-2\sum\limits_{i=x,y}\alpha_{i}\gamma_{i}k_{i}^{2}\right) }{\left( \mathbb{I}_{1}\left(\boldsymbol{k},\boldsymbol{q},\boldsymbol{\alpha},\mu  \right) +\left(\mathbb{I}_{2}^{2}\left(\boldsymbol{k},\boldsymbol{q},\boldsymbol{\alpha} \right)+\left(\chi q_{y} \right)^{2}  \right)^{1/2} \right)^{2} }\right) \left\langle \sigma_{1}\right\rangle -\left(\chi k_{y} \right) \left\langle \sigma_{2}\right\rangle \notag \\
&+&  \sum\limits_{\boldsymbol{q} \mu}\left(  \frac{\left|\mathcal{M}_{\mu \boldsymbol{q}} \right|^{2} \left(\chi k_{y}\right) }{\left( \mathbb{I}_{1}\left(\boldsymbol{k},\boldsymbol{q},\boldsymbol{\alpha},\mu  \right) +\left(\mathbb{I}_{2}^{2}\left(\boldsymbol{k},\boldsymbol{q},\boldsymbol{\alpha}\right)+\left(\chi q_{y} \right)^{2}  \right)^{1/2} \right)^{2} }\right) \left\langle \sigma_{2}\right\rangle, \qquad \left\langle \sigma_{i} \right\rangle=\left\langle \Psi_{\lambda} \right| \sigma_{i} \left| \Psi_{\lambda}\right\rangle . \label{10}
\end{eqnarray}

Here, $  \left| \Psi_{\lambda}\right\rangle  $ are the eigenvectors of the unperturbed $ \boldsymbol{k}\cdot\boldsymbol{p} $ Hamiltonian [Eq.(\ref{2})], which can be expressed as,

\begin{eqnarray}
 \left| \Psi_{\lambda}\right\rangle   &=&\frac{1}{\sqrt{2}}\left[ 
\begin{array}{c}
N_{\lambda} \\ \\
1
\end{array}%
\right], \qquad N_{\lambda}=\dfrac{E_{\lambda}^{0}-\left(u_{0}+\sum\limits_{i=x,y}\eta_{i}k_{i}^{2} \right) }{\left(\delta+\sum\limits_{i=x,y}\gamma_{i}k_{i}^{2} \right)}, \notag
\end{eqnarray}
which corresponds to the following eigenvalues for electron ($ \lambda=1 $) and holes ($ \lambda=-1 $)
\begin{equation}
E_{\lambda}^{0}=u_{0}+\sum\limits_{i=x,y}\eta_{i}k_{i}^{2}+\lambda\left[\left( \delta+\sum\limits_{i=x,y}\gamma_{i}k_{i}^{2}\right)^{2} +\left( \chi k_{y}\right)^{2} \right]^{1/2} \label{11}
\end{equation}

It can be seen that, the energy expression of pristine MBP in Eq.(\ref{11}) leads to a band gap, i.e., $ E_{g}=1.52 $ eV. Finally, the eigenvalues given by Eq.(\ref{10}) can be represented as,
\begin{eqnarray}
E_{\lambda}&=&E_{\lambda}^{0}+\sum\limits_{i=x,y}\left( \alpha_{i}^{2}\eta_{i}k_{i}^{2}+\alpha_{i}^{2}\gamma_{i}k_{i}^{2}\right)+\mathbf{\Sigma}_{1}\left(\boldsymbol{k},\boldsymbol{\alpha} \right) \notag \\
&+&\dfrac{\lambda}{\left[ \left(\delta+ \sum\limits_{i=x,y}\gamma_{i}k_{i}^{2}\right)^{2}+\left(\chi k_{y} \right)^{2}  \right]^{1/2} }\left[\sum\limits_{i=x,y}\left(\delta+ \alpha_{i}^{2}\gamma_{i}k_{i}^{2}\right) \mathbf{\Sigma}_{2}\left(\boldsymbol{k},\boldsymbol{\alpha} \right)-\left( \chi k_{y}\right) \mathbf{\Sigma}_{3}\left(\boldsymbol{k},\boldsymbol{\alpha} \right)  \right], \label{12}  
\end{eqnarray}
with
\begin{eqnarray}
\mathbf{\Sigma}_{1}\left(\boldsymbol{k},\boldsymbol{\alpha} \right)&=&\sum\limits_{\boldsymbol{q} \mu} \left( \frac{-2\left|\mathcal{M}_{\mu \boldsymbol{q}} \right|^{2} }{\mathbb{I}_{1}\left(\boldsymbol{k},\boldsymbol{q},\boldsymbol{\alpha},\mu  \right) +\left(\mathbb{I}_{2}^{2}\left(\boldsymbol{k},\boldsymbol{q},\boldsymbol{\alpha}\right)+\left(\chi q_{y} \right)^{2}  \right)^{1/2} }\right)  \notag \\
&+&\sum\limits_{\boldsymbol{q} \mu}\left(  \frac{\left|\mathcal{M}_{\mu \boldsymbol{q}} \right|^{2} \left( \mathbb{I}_{1}\left(\boldsymbol{k},\boldsymbol{q},\boldsymbol{\alpha},\mu  \right)-2\sum\limits_{i=x,y}\alpha_{i}\eta_{i}k_{i}^{2}\right) }{\left( \mathbb{I}_{1}\left(\boldsymbol{k},\boldsymbol{q},\boldsymbol{\alpha},\mu  \right) +\left(\mathbb{I}_{2}^{2}\left(\boldsymbol{k},\boldsymbol{q},\boldsymbol{\alpha} \right)+\left(\chi q_{y} \right)^{2}  \right)^{1/2} \right)^{2} }\right)  \notag \\
\mathbf{\Sigma}_{2}\left(\boldsymbol{k},\boldsymbol{\alpha} \right)&=& \sum\limits_{\boldsymbol{q} \mu}\left(  \frac{\left|\mathcal{M}_{\mu \boldsymbol{q}} \right|^{2} \left( \mathbb{I}_{2}\left(\boldsymbol{k},\boldsymbol{q},\boldsymbol{\alpha}\right)-2\sum\limits_{i=x,y}\alpha_{i}\gamma_{i}k_{i}^{2}\right) }{\left( \mathbb{I}_{1}\left(\boldsymbol{k},\boldsymbol{q},\boldsymbol{\alpha},\mu  \right) +\left(\mathbb{I}_{2}^{2}\left(\boldsymbol{k},\boldsymbol{q},\boldsymbol{\alpha}\right)+\left(\chi q_{y} \right)^{2}  \right)^{1/2} \right)^{2} }\right) \notag \\
\mathbf{\Sigma}_{3}\left(\boldsymbol{k},\boldsymbol{\alpha} \right)&=&\sum\limits_{\boldsymbol{q} \mu}\left(  \frac{\left|\mathcal{M}_{\mu \boldsymbol{q}} \right|^{2} \left(\chi k_{y}\right) }{\left( \mathbb{I}_{1}\left(\boldsymbol{k},\boldsymbol{q},\boldsymbol{\alpha},\mu  \right) +\left(\mathbb{I}_{2}^{2}\left(\boldsymbol{k},\boldsymbol{q},\boldsymbol{\alpha} \right)+\left(\chi q_{y} \right)^{2}  \right)^{1/2} \right)^{2} }\right). \label{13}
\end{eqnarray} 

Summations in Eq.(\ref{13}) can be replaced with integral over $ \boldsymbol{q} $, i.e., $ \sum_{\boldsymbol{q}}\rightarrow \left( A/4\pi^{2}\right) \int \int dq_{x} dq_{y}$, where $ A $ is the surface area of MBP. While taking the integrals we introduce a cut-off phonon wavevector, $ q_{cut} $ . The eigenvalues are calculated by numeric minimization of Eq.(\ref{12}) with respect to the two different variation parameters, $ \alpha_{x} $ and $ \alpha_{y} $, which is necessary to take into account the anisotropic nature of MBP along the $x $ and $y$ direction, respectively. 

\section{Results and Discussion}
Due to the polaronic effect caused by a polar substrate, the energy band gap of MBP is increased depending on the interlayer distance, dielectric constants, and characteristic phonon frequencies of a substrate. Indeed, from Eq.(\ref{12}) one can see that, while the second and third terms cause a uniform shift of the valance and conduction bands and do not produce any gap change, the fourth term changes the positions of the bands for $\lambda=-1$ and $\lambda=+1$ individually thus inducing a gap broadening ($ \Delta E_{g}$). Although electron-phonon coupling in MBP due to the surface optical phonon modes of a substrate can be regarded as weak, the preferred direction choice made in Eq.(\ref{8}) allows us to expect more accurate eigenvalues of the total Hamiltonian [Eq.(\ref{4})] by taking into account correlations introduced by electron recoil \cite{lee1953motion}.

\begin{figure} [h!]
	\centering
	\includegraphics[width=0.95\textwidth]{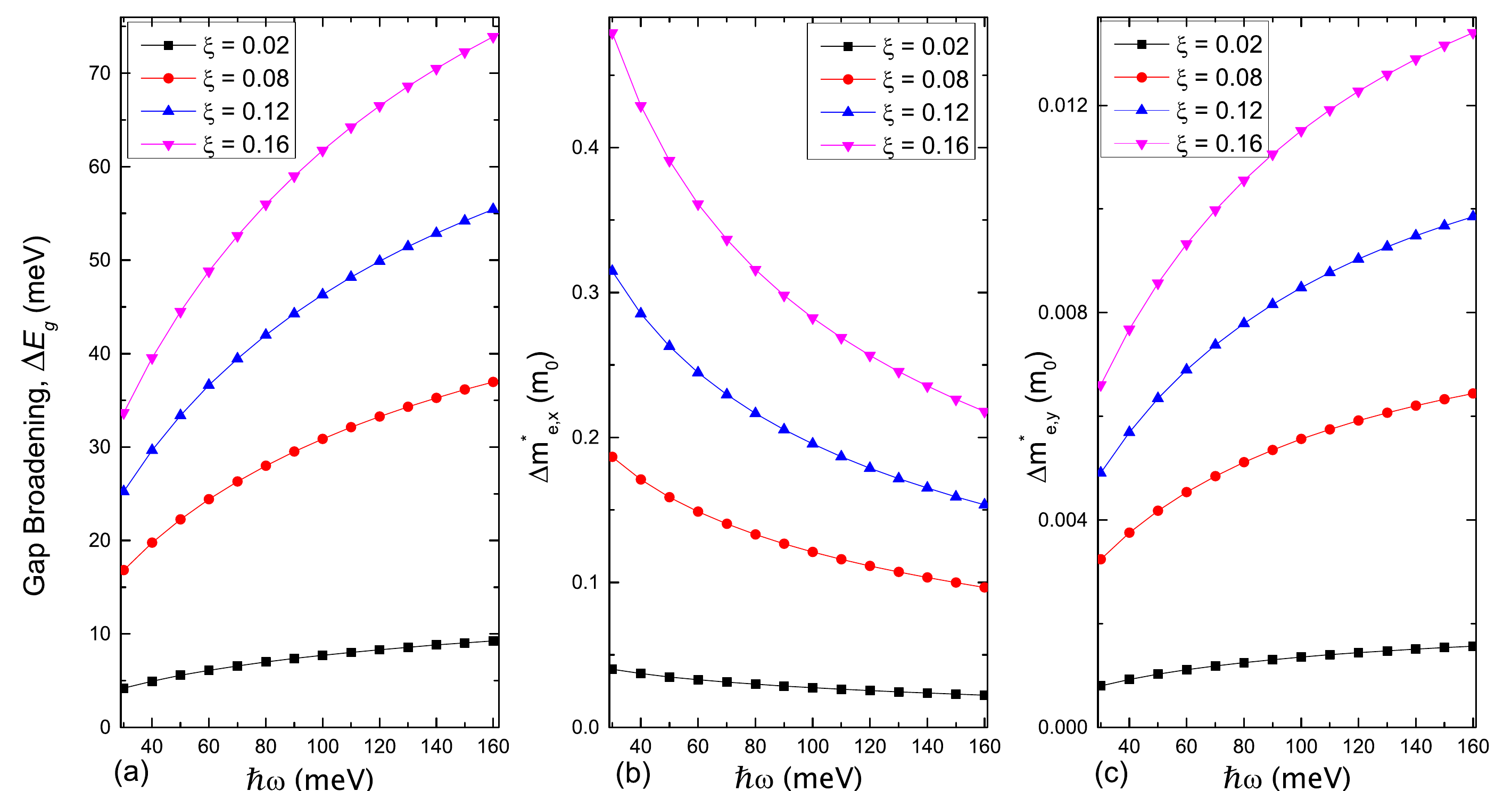}
	
	\caption{(a) Gap broadening ($ \Delta E_{g}$) and (b),(c) renormalization of direction-dependent effective masses in MBP, as a function of phonon frequency, $ \hbar \omega $, for different effective screening constants, $ \xi$. Each case corresponds to a fixed MBP-substrate separation, $z=2 $ \AA{}.}
	\label{FIGURE2}
\end{figure}

We first analyze general dependence of the energy gap and effective masses in MBP on the characteristic substrate frequency, $\omega $ and on its effective screening parameter, $\xi $. For simplicity, we take $ \omega=\omega_{1}=\omega_{2} $. The corresponding dependencies are shown in Fig.(\ref{FIGURE2}). As can be seen from Fig.(\ref{FIGURE2}) (a), the gap increases with increasing the values of both $ \hbar \omega $ and $\xi $. From the Figs.(\ref{FIGURE2}) (b) and (c), it can be also seen that for higher $ \hbar \omega $, effective masses are decreasing along the $ x $ direction, but increasing along the $ y $ direction. The qualitative difference of the phonon frequency dependence of the masses along the $ x $ and $ y $ directions arises from the linear $ \chi k_{y}$ term which appears in Eq.(\ref{11}). Note that, the anisotropy of MBP not only originates from the coefficients in $ k_{x}^{2} $ and $ k_{y}^{2} $ terms, i.e., $ \eta $ and $ \gamma $, but also from the linear $ k_{y} $ term in the Hamiltonian Eq.(\ref{2}). In other words, charge carriers of MBP show both Schr{\"{o}}dinger-like and Dirac-like character in addition to the anisotropy of the MBP spectrum \cite{1504.02452v1}. Therefore, when the electron-phonon interaction is switched on, the dynamics of charge carriers also changes significantly along both directions in addition to the existing anisotropy of MPB. 

\begin{figure} [h!]
    \centering
   
    \includegraphics[width=0.70\textwidth]{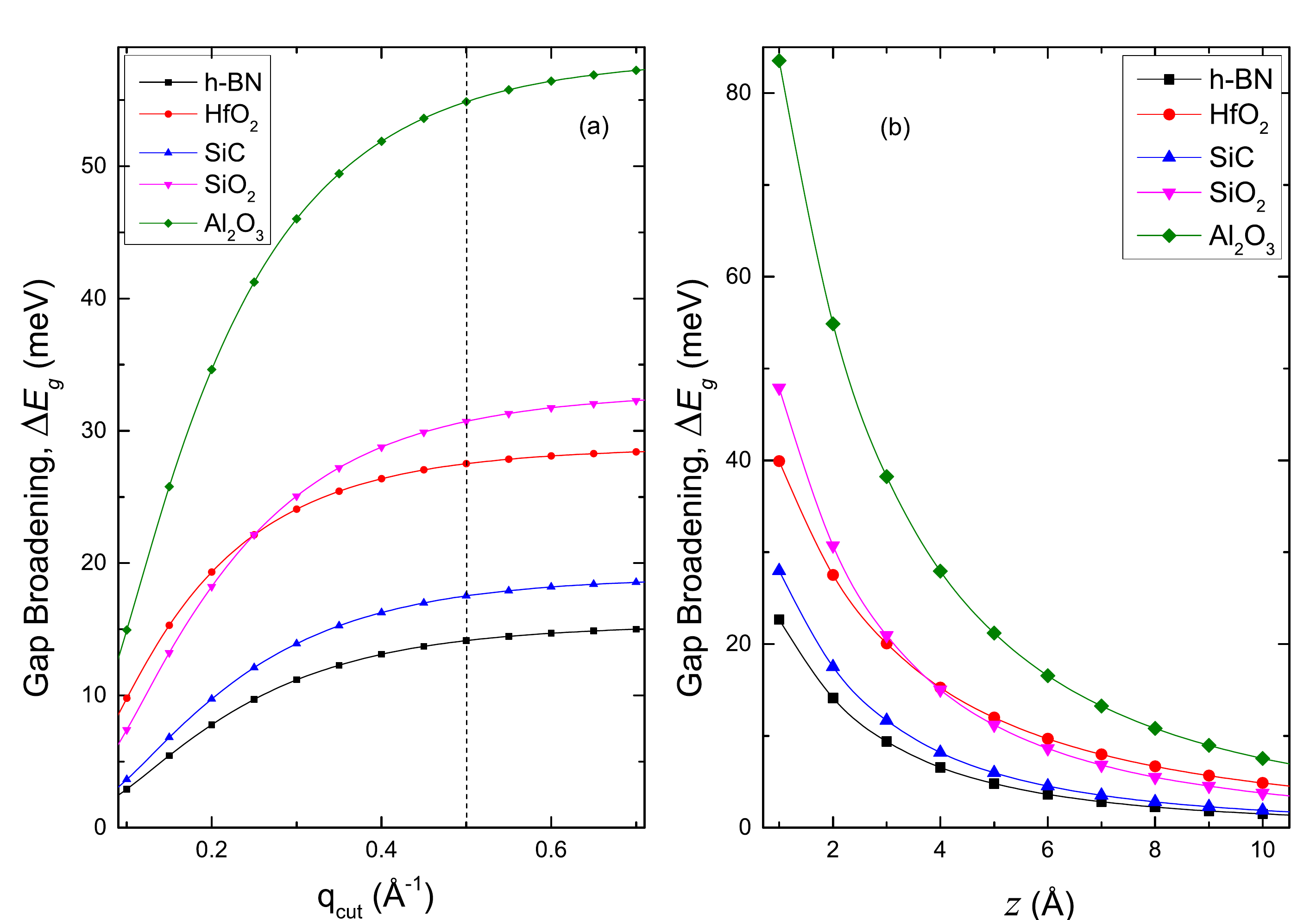}
    \caption{Evolution of a band gap in MBP with the cutoff wavevector $q_{cut} $ (a) and with distance to the substrate (b) shown for $ z=2 $ \AA{}. Different curves correspond to different substrates. Dashed line corresponds to $q_{cut} $ used in this paper for other calculations.}
    \label{FIGURE3}
\end{figure}

In practical calculations, surface optical phonon modes of polar substrate emerge from its finite size and they can be extracted from its bulk transverse optical phonon modes ($\omega_{TO_{\mu}}$) using the formula, $ \hbar \omega_{\mu}=\hbar \omega_{TO_{\mu}} F_{\mu} $, $ F_{1}=\left[\left(\kappa_{0}+1 \right) /\left(\kappa_{int}+1 \right) \right]^{1/2} $ and $ F_{2}=\left[\left(\kappa_{int}+1 \right) /\left(\kappa_{\infty}+1 \right) \right]^{1/2} $ \cite{Wang:1972aa,fischetti2001effective,Tan2013aa}. Here, $ \kappa_{int} $ is the intermediate frequency dielectric constant of the substrate. Surface optical phonon modes and dielectric constants of polar substrates are given in Table \ref{tab1}, which have been taken from Refs. \onlinecite{konar2010effect,wang2015energy}. Among these substrates, h-BN has the highest surface optical phonon frequencies, whereas Al$ _{2} $O$ _{3} $ has the highest effective screening constant. In terms of practical applications, polar substrates might give rise to considerable modifications of the electronic band structure of supported samples, which are expected to be especially important for scattering processes. To examine the effect of specific substrates on $ \Delta E_{g}$ in MBP, we have considered a series of typical polar substrates: h-BN, SiC, SiO$_{2}$, HfO$_{2}$, Al$_{2}$O$_{3}$. One can see from Fig.(\ref{FIGURE3})(b), $ \Delta E_{g}$ decreases with $ z $ and converged to the band gap of pristine MBP after some values of $ z $. It is also clear from the Fig.(\ref{FIGURE3}) (b) that, while the influence of the effective screening constant, $ \xi $, is more pronounced for higher $ z $, the phonon frequencies become more effective at lower $ z $. Since $ \xi $ values of SiO$ _{2} $ is lower than HfO$ _{2}$, $ \Delta E_{g}$ of SiO$ _{2} $ higher than HfO$ _{2} $ below $ z\approx5 $  \AA{}, which is related to the relatively higher phonon frequencies of SiO$ _{2} $. $ \Delta E_{g}$ is also dependent on the cut-off wavevector $ q_{cut}$, which converges at around $ 0.5 \sim 0.6 $ \AA{}$ ^{-1} $. The effect of $ q_{cut} $ on $ \Delta E_{g}$ becomes insignificant when $ z $ increases, and means that only the phonons with low $ q $, are strongly coupled with the charge carriers in MBP. We considered $ q_{cut}=0.5 $ \AA{}$ ^{-1} $ for all numerical calculations presented in this paper.
 \begin{table} [h]
 	 \begin{center}
 	 		\caption{Surface-optical phonon modes, dielectric constants, and effective screening constants of the polar substrates used in this paper. Parameters have been taken from Refs. \onlinecite{konar2010effect,wang2015energy}.} \label{tab1}
 	 	\begin{tabular}{l*{6}{c}}
 	 		Substrate  &\qquad $ \hbar \omega_{1} $ (meV) & \quad$ \hbar \omega_{2} $ (meV)& \quad$ \kappa_{0} $ &\quad $ \kappa_{\infty} $ & \quad $ \xi$ &\\
 	 		\hline
 	 		h-BN           & 101 & 195 & \quad5.1 & \quad4.1 & \quad0.03 & \\
 	 		SiC            & 167 & 116 & \quad9.7 & \quad6.5 & \quad0.04 &  \\
 	 		SiO$ _{2} $    & 146 & 60 & \quad3.9 & \quad2.5 & \quad0.08 &  \\
 	 		HfO$ _{2} $    & 53 & 19 & \quad22.0 & \quad5.0 & \quad0.12 &  \\
 	 		Al$ _{2} $O$ _{3} $    & 94 & 55 & \quad12.5 & \quad3.2 & \quad0.16 & \\
 	 	\end{tabular}
 	 \end{center}
 \end{table}
 
\begin{figure} [h!]
    \centering
        \includegraphics[width=\textwidth]{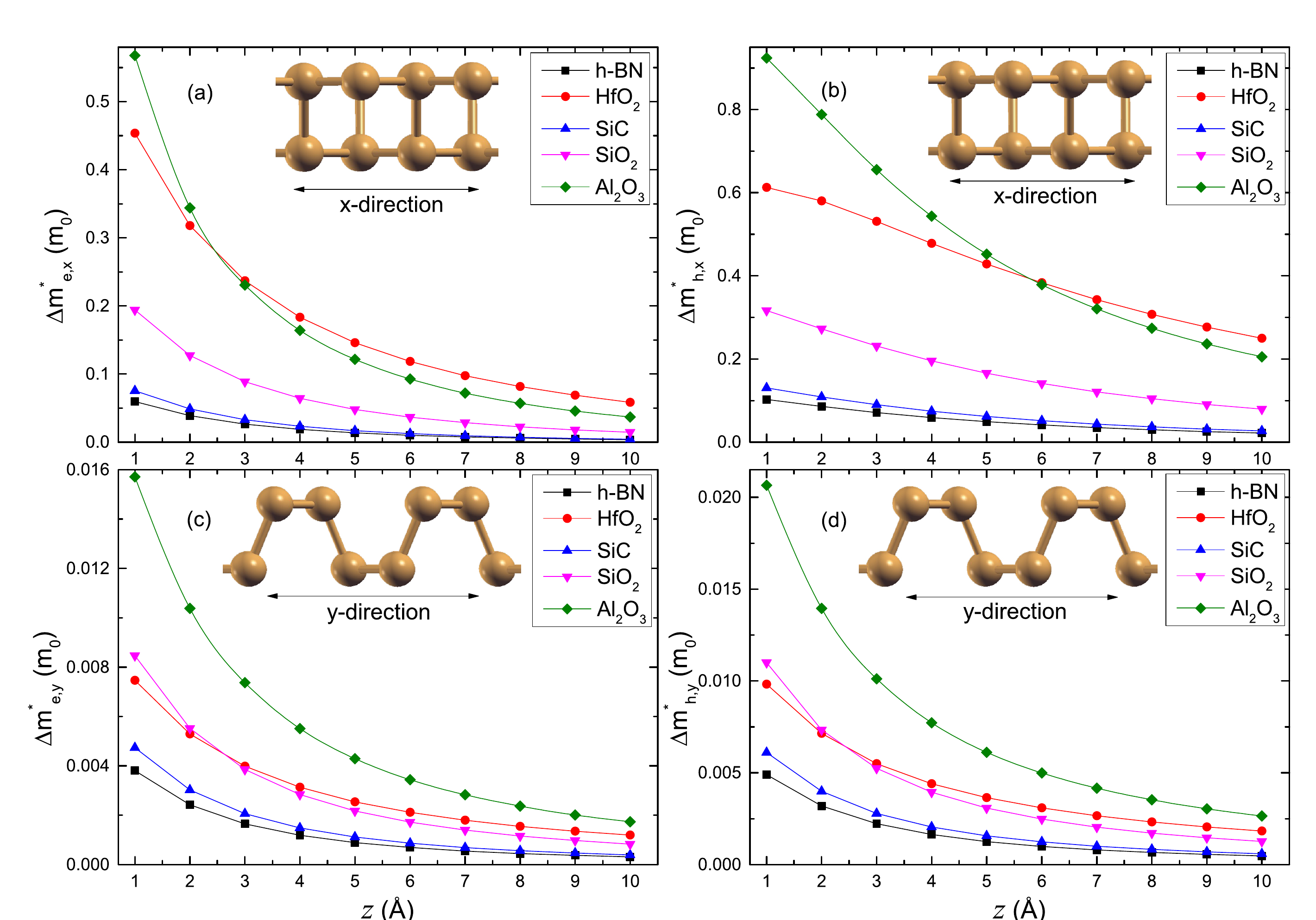}
    
    \caption{Renormalization of effective masses in MBP due to the polaronic effect induced by different substrates shown as a function of the MBP-substrate separation. The values are given in units of free electron mass, $ m_{0}$.}
    \label{FIGURE4}
\end{figure}

 As a next step, we analyze the effective masses. The effective masses can be calculated as, $ m_{\lambda,i}^{*}=\hbar^{2}/\left(\partial ^{2}E_{\lambda}/\partial k_{i}^{2} \right)  $, where $ i=x,y $ and $ \lambda=1 $($ -1 $) corresponds to electron (hole) states. In pristine MBP, charge carriers have different effective masses due to the anisotropy of the system along the $ x $ and $ y $ direction. While the $ x $ component of effective masses is, $ m_{\lambda,x}^{*}=\hbar^2/2\left(\eta_{x}+\lambda\gamma_{x} \right)  $ ($ m_{e,x}^{*}=0.846 $ $ m_{0} $ and $ m_{h,x}^{*}=1.14 $$ m_{0}) $, the $ y $ component can be found by expanding the expression around $ k_{y}=0 $ point as,  $ m_{\lambda,2}^{*}=\hbar^2/2\left(\eta_{x}+\lambda\gamma_{x}+\lambda \left( \chi^{2}/2\delta \right) \right)  $ ($ m_{e,y}^{*}=0.166 $ $ m_{0} $ and $ m_{h,y}^{*}=0.182 $ $ m_{0} $, where $ m_{0} $ is the free electron mass). The dressed effective masses can be found by following the same procedure. In this case, not only $ y $ components but also $ x $ components of the effective masses can be obtained by expanding around $ k_{x}=0 $ point. The many-body correction to effective masses ($ \Delta m_{\lambda,i}^{*} $=$ \bar{m}_{\lambda,i}^{*}-m_{\lambda,i}^{*} $) of the charge carriers in MBP is presented in Fig.(\ref{FIGURE4}) for different $ z $ values, where $ \bar{m}_{\lambda,i}^{*} $ are the dressed effective masses under the influence of electron-phonon interaction. The effective masses of the charge carriers of MBP increase by the influence of the polaronic effect. In Fig.(\ref{FIGURE4})(a),and (b), the surface polaronic effect on the electron and hole effective masses along the $ x $ direction is presented. One can see that along the $ x$-direction hole-phonon has stronger coupling than the electron-phonon one. Since the effective masses along the $ x $ direction decrease with $ \hbar \omega $ as it shows in Fig.\ref{FIGURE2}(b), HfO$ _{2} $-induced effective mass enhancement is higher compared to other substrates at high $ z $ values. However, the effect of $ \xi $ becomes more dominant at low $ z $. In the vicinity of $ z=2 $ \AA{} and $ z=6 $ \AA{}, it is clear from Fig.(\ref{FIGURE4})(a),and (b) Al$ _{2} $O$ _{3} $ has the largest effect for both electrons and holes due to the higher effective screening constant. In Fig.(\ref{FIGURE4})(c), and (d), we show the renormalization of the electron and hole masses along the $ y $ direction which is substantially smaller compared to the $ x $ components. This can be related with a weaker electron-phonon coupling along the $ y $ direction. In contrast to the effective masses along the $ x $ direction, the $ y $ component of effective masses increases with $ \hbar \omega $ as it is seen from Fig.\ref{FIGURE2}(c). Therefore, the $\Delta m_{\lambda,y}^{*} $ show similar behavior to $ \Delta E_{g}$
 given in Fig.(\ref{FIGURE3}). That is, the effective screening constant, $ \xi $ basically determines the behavior of effective masses along the $ y $ direction at all distances, whereas the effects of phonon frequencies on effective masses become more pronounced at low $ z $.   

\section{Conclusion}
In summary, we have shown that electron-phonon interaction in MBP on polar substrates gives rise to a polaronic effect, which is an efficient many-body mechanism affecting the band properties of the system. We have used the LLP method to capture the effects of electron-phonon coupling in the energy spectrum of MBP within the long-wavelength approximation. In contrast to standard perturbation theory which is valid only in the weak coupling regime, this method is also applicable to the intermediate interactions. In comparison to the original formulation of the LLP method, we have considered two different variation parameters, i.e., $ \alpha_{x} $ and $ \alpha_{y} $, corresponding to the $ x $ and $ y $ direction, respectively, which is governed by the anisotropy of effective masses in MBP. Following this method, we have calculated the band gap $ \Delta E_{g}$ and effective mass variations in MBP arising from the coupling of carriers with surface optical phonon modes of different substrates. We have found that, the highest contribution to $ \Delta E_{g}$ comes from the Al$ _{2} $O$ _{3}$ substrate. We have also shown that, while the effective screening constant is the main parameter that determines the gap variations at all interlayer separations, phonon energies of a substrate are important only at low $ z $. At $ z=2.5$ \AA{}, which is a typical interlayer distance between graphene and SiO$_{2}$ \cite{rudenko2011interfacial}, the largest $ \Delta E_{g}$ is the order of $ 50$ meV. Our analysis of the effective mass renormalization shows that the interaction of charge-carrier coupling with phonons is significantly stronger along the $ x $ direction than the $ y $ direction. Due to its large effective screening constant, Al$ _{2} $O$ _{3}$-induced mass enhancements are higher with respect to the other substrates at low $ z $. For higher $ z $, one should also involve the energies of surface optical phonons to understand the behavior of effective mass renormalization. We have also shown that the dependence of effective masses on $ \hbar \omega$ is qualitatively different for armchair and zigzag directions. This is due to the highly anisotropic nature of the continuum Hamiltonian of MBP. Moreover, it should be noted that the carrier-surface optical phonon coupling increases the effective mass anisotropy of MBP.  
Our results provide a starting point for the experimental verification of the surface polaronic effect in supported or encapsulated MBP. We also believe that our results will lead to deeper understanding of the charge-carrier dynamics in MBP from both theoretical and experimental perspectives.
\begin{acknowledgments}
A.M. and Y.M. acknowledge "The Scientific and Technological Research Council of Turkey (TUBITAK)" through "BIDEB-2219 Postdoctoral Research Fellowship.". Funding from the European Union Seventh Framework Programme under Grant Agreement No. 604391 Graphene Flagship is also acknowledged.
\end{acknowledgments}

\section*{References}

\bibliography{referans}
\bibliographystyle{apsrev4-1.bst}

\end{document}